\documentclass[journal]{IEEEtran}
\pdfoutput=1
\ifCLASSINFOpdf
\else
   \usepackage[dvips]{graphicx}
\fi
\usepackage{url}
\usepackage{cite}
\usepackage{graphicx}
\usepackage{booktabs}
\usepackage{float}
\usepackage{amsmath}
\usepackage{alphabeta}
\usepackage{algorithm}  
\usepackage{algorithmic}
\usepackage{multirow}
\usepackage{makecell}
\usepackage{array}
\begin{document}
\newcolumntype{L}[1]{>{\raggedright\let\newline\\\arraybackslash\hspace{0pt}}m{#1}}
\newcolumntype{C}[1]{>{\centering\let\newline\\\arraybackslash\hspace{0pt}}m{#1}}
\newcolumntype{R}[1]{>{\raggedleft\let\newline\\\arraybackslash\hspace{0pt}}m{#1}}
\title{MarkNerf:Watermarking for Neural Radiance Field}

\author{Lifeng Chen,\IEEEmembership{} Jia Liu, Yan Ke, Wenquan Sun, Weina Dong and Xiaozhong Pan\IEEEmembership{}
\thanks{Manuscript received 31 August 2023; accepted X September 2023. Date of publication X X 2023. This work was supported in part by the General Program of the National Natural Science Foundation of China under Grant 62272478.The associate editor coordinating the review of this manuscript and approving it for publication was XXX (Corresponding author: Jia Liu.)}
\thanks{The authors are with the College of Cryptography Engineering, Engineering University of PAP, Xi’an Shaanxi 710086, China; Key Laboratory of Network and Information Security of PAP (Engineering University of PAP), Xi’an Shaanxi 710086, China. (e-mail:703792110@qq.com;liujia1022@gmail.com;
3011745933@qq.com;18691481090@qq.com;13609181367@139.com).}}

\markboth{Journal of \LaTeX\ Class Files, Vol. 14, No. 8, August 2015}
{Shell \MakeLowercase{\textit{et al.}}: Bare Demo of IEEEtran.cls for IEEE Journals}
\maketitle

\begin{abstract}
A watermarking algorithm is proposed in this paper to address the copyright protection issue of implicit 3D models. The algorithm involves embedding watermarks into the images in the training set through an embedding network, and subsequently utilizing the NeRF model for 3D modeling. A copyright verifier is employed to generate a backdoor image by providing a secret perspective as input to the neural radiation field. Subsequently, a watermark extractor is devised using the hyperparameterization method of the neural network to extract the embedded watermark image from that perspective. In a black box scenario, if there is a suspicion that the 3D model has been used without authorization, the verifier can extract watermarks from a secret perspective to verify network copyright. Experimental results demonstrate that the proposed algorithm effectively safeguards the copyright of 3D models. Furthermore, the extracted watermarks exhibit favorable visual effects and demonstrate robust resistance against various types of noise attacks.
\end{abstract}


\IEEEpeerreviewmaketitle

\section{INTRODUCTION}
\IEEEPARstart{N}{eural} Radiation Field (NeRF) [1] is a deep learning model for modeling three-dimensional implicit spaces [2]. It uses neural networks to learn a continuous function that maps spatial coordinates to density and color, achieving the synthesis of new perspectives. 3D rendering technology based on implicit representation has become one of the most popular fields of computer vision research. It can be predicted that in the near future, people will share 3D content online, just like sharing 2D images and videos, and protecting implicit 3D data will become an important issue in the future.

Before NeRF, 3D data was mostly represented in the form of point clouds, voxels, or triangular meshes. The methods for copyright protection of these forms of 3D data were usually divided into three categories: embedding watermarks directly by translating, rotating, scaling 3D shapes [3-8], modifying 3D model parameters to embed watermarks [9-12], and using deep learning techniques to embed watermarks [13-16]. However, unlike traditional 3D models, NeRF is a function that represents the color and density of each point in a 3D scene. There is no specific visible data such as the 3D model and 3D model parameters, so it is not possible to translate, rotate, or embed watermarks in 3D model parameters. Moreover, the neural radiation field is trained through the MLP network structure, and adjusting the network structure to embed watermarks can affect the effectiveness of 3D rendering.

In 2022, Li et al.[17] established the connection between information hiding and NeRF for the first time and proposed the StegaNeRF scheme for information hiding. This scheme first trains a standard NeRF model, and then treats it as a new perspective generator. While training the message extraction network, the NeRF network is retrained to ensure that the 2D images rendered from the new StegaNeRF network can accurately extract messages.

In this letter, we propose a watermarking scheme for NeRF, which first embeds watermarks into the images in the training set through an embedding network. Next, use the NeRF model for 3D model modeling. The copyright verifier generates an image from a secret perspective as the input of the neural radiation field, and then uses the hyperparameterization method of the neural network to design a watermark extractor to obtain the embedded watermark from that perspective. In the black box scenario [18-25], once it is suspected that the 3D model has been used without authorization, the verifier can extract the watermark from a secret perspective to verify network copyright. Compared with StegaNeRF, this method does not require secondary training of NeRF, ensuring the quality of 3D model rendered images.

This letter made the following contributions:

\quad \textbullet\quad We have used parameterized methods to extract watermarks from rendered images from specific perspectives by training a simple network.

\quad \textbullet\quad We use the perspective input of the NeRF model and use the secret perspective as the key information for watermark extraction. Due to the continuity of the new perspective synthesis and the huge key space, the security of the watermark information is guaranteed.

\quad \textbullet\quad We simulated the noise processing of implicit expression networks by introducing a noise layer on the training image, making the method robust.
\section{PROPOSED METHOD}
\begin{table}
\label{table}
\setlength{\tabcolsep}{5pt}
\begin{tabular}{p{225pt}}
\hline
\textbf{Algorithm 1}  Typical application scenario of MarkNeRF
\\ \hline
1: Alice acquires a set of images depicting the 3D scene.\\
2: Alice embeds a watermark within the images and generates the NeRF model N.\\
3: Alice publicly shares model N on the Internet to allow others to enjoy the 3D scene.
4: Without obtaining Alice’s permission, Bob acquires model N and uploads it to the network under his own name.\\
5: Upon discovering the publication of model N by Bob, Alice employs the secret perspective m to extract the watermark. This serves as evidence to verify that Alice is the rightful owner of model N, rather than Bob.\\
6: Bob’s unauthorized publication constitutes copyright infringement, necessitating the withdrawal of the infringing content.\\
\hline
\end{tabular}
\label{tab1}
\end{table}

\begin{figure*}
\centerline{\includegraphics[width=1.0\textwidth  ]{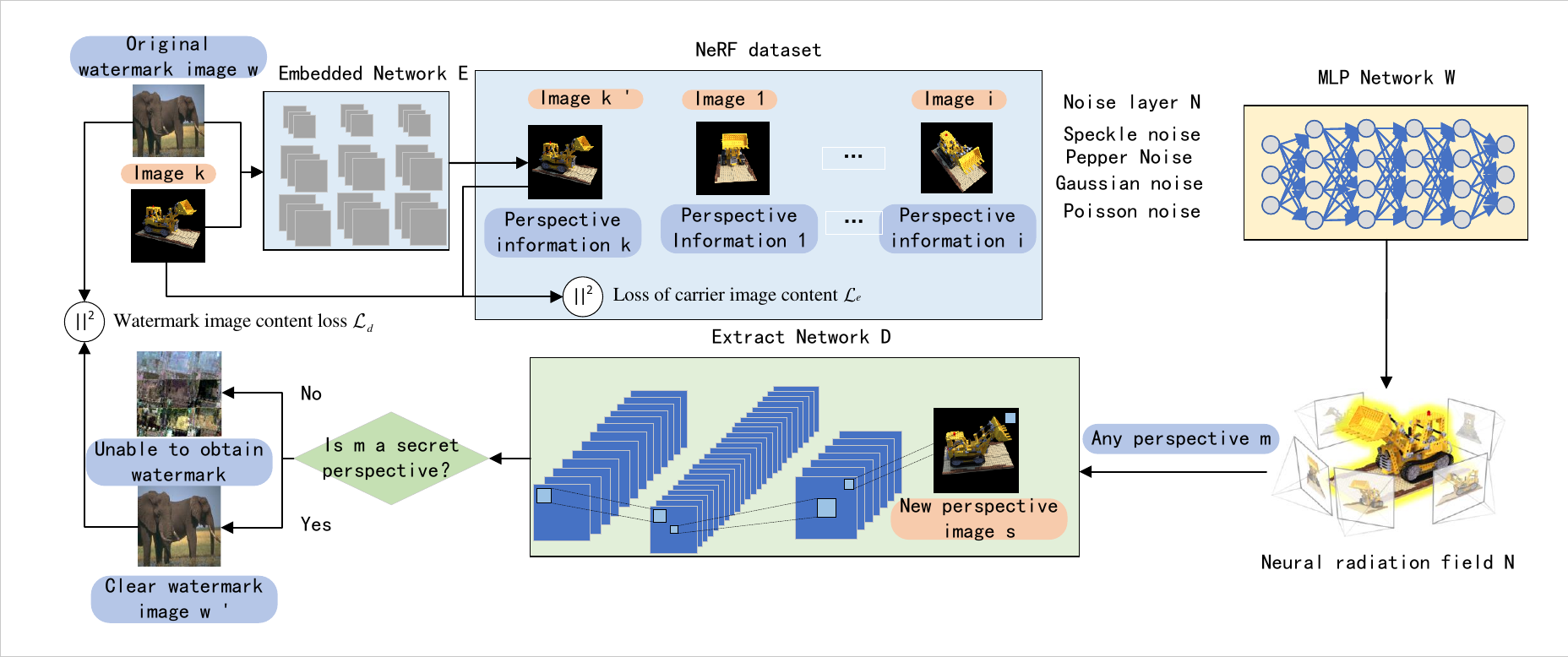}}
\caption{Model Overview. The embedding network E takes the watermark image w and an image k from the NeRF training dataset (taking the Lego dataset as an example) as inputs, outputs the image k 'with embedded watermark information, and then processes the images 1 to i from the Lego dataset with noise layer processing and corresponding perspective information to input into the MLP network M to train the neural radiation field N. Then, a parameterized method is used to train an extraction network D, Finally, give the neural radiation field N an arbitrary perspective as input to generate a new perspective image s. If the perspective m is the secret perspective used during training, a clear watermark image w 'can be extracted, otherwise the watermark image cannot be extracted.}
\end{figure*}

The framework proposed in this article is shown in Figure 1, which mainly includes four stages: (1) embedding watermark, (2) adding noise, (3) training NeRF model, and (4) extracting watermark.
\begin{figure}
\centerline{\includegraphics[width=\columnwidth]{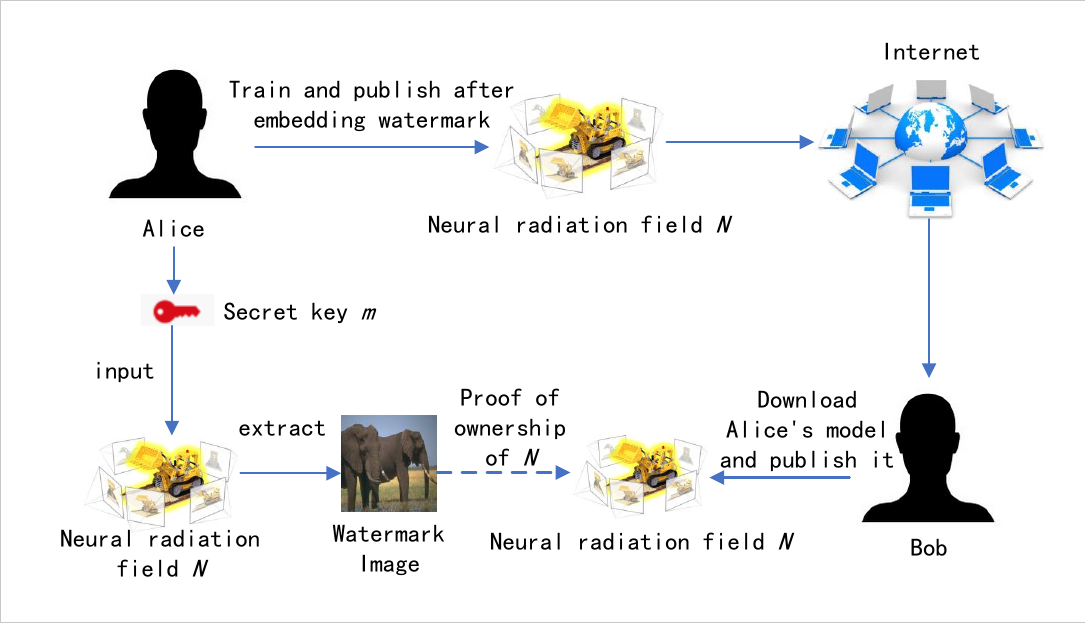}}
\caption{Application Scenario Process.}
\end{figure}

\subsection{Embedded network}The embedded network, denoted as E, takes the watermark information with the size of 3×W×H and the image k from the Lego dataset as inputs. The watermark image w is then passed through a Conv layer with an output channel of 32, resulting in the feature map A with the size of 32×W×H. Furthermore, the image k is fed through another Conv layer with an output channel of 32 to obtain the feature map B, also with the size of 32×W×H.
\begin{equation}
\begin{array}{l}
\left\{ {\begin{array}{*{20}{c}}
{A = Con{v_{3 -  > 32}}({\rm{a}})}\\
{B = Con{v_{3 -  > 32}}({\rm{k}})}\\
{C = Con{v_{32 + 32 -  > 32}}(Cat(A,B))}\\
{D = Con{v_{32 + 32 + 32 -  > 32}}(Cat(A,B,C))}\\
{{\rm{k'}} = Con{v_{32 + 32 + 32 + 32 -  > 3}}(Cat(A,B,C,D))}
\end{array}} \right.
\end{array}
\end{equation}
Within the embedded network, we utilize a convolutional kernel with a size of 3×3, a step size of 1, and padding of 1. The ReLU activation function is employed, and a BN (batch normalization) layer is applied to normalize the data. The specific architecture of the embedded network is illustrated in Figure 4.
\begin{figure}
\centerline{\includegraphics[width=\columnwidth]{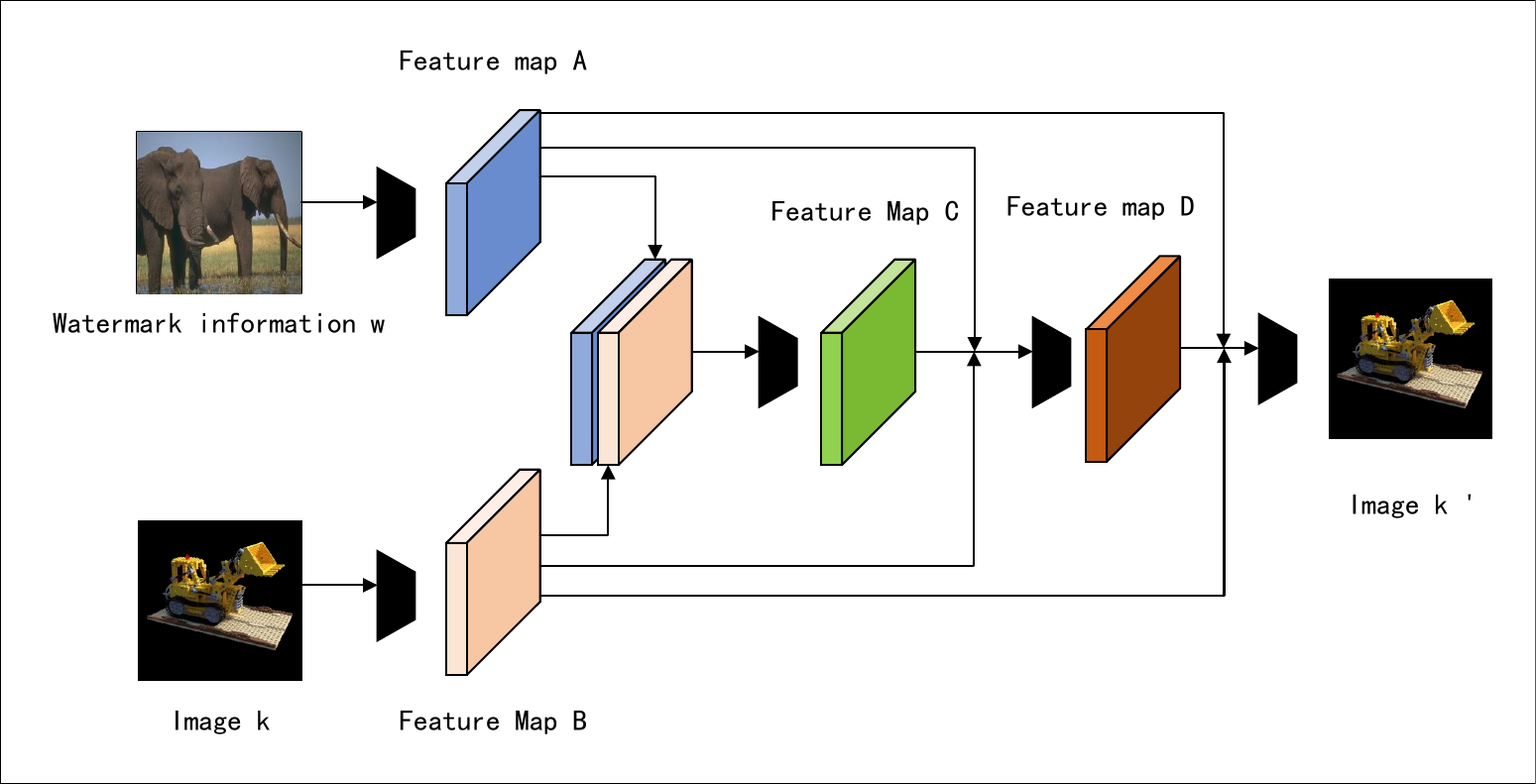}}
\caption{Watermark embedding network structure.}
\end{figure}
\subsection{Noise layer} Upon exposure to noise, the neural radiation field may result in image distortion under the secret view [1]. To enhance the robustness of the proposed method, a noise layer is introduced for model processing. This study posits two approaches to incorporate noise into the implicitly represented 3D model: one involves adding noise to the network model parameters themselves, as demonstrated in reference [1], while the other entails introducing noise to the input data to simulate alterations in the network parameters themselves. For the purpose of simulating NeRF’s susceptibility to noise, this study adopts the latter approach. Specifically, Gaussian noise, salt and pepper noise, speckle noise, and Poisson noise are utilized to emulate various types of distortion. Refer to Figure 5 for the visual representation of the noise image.
\begin{figure}
\centerline{\includegraphics[width=\columnwidth]{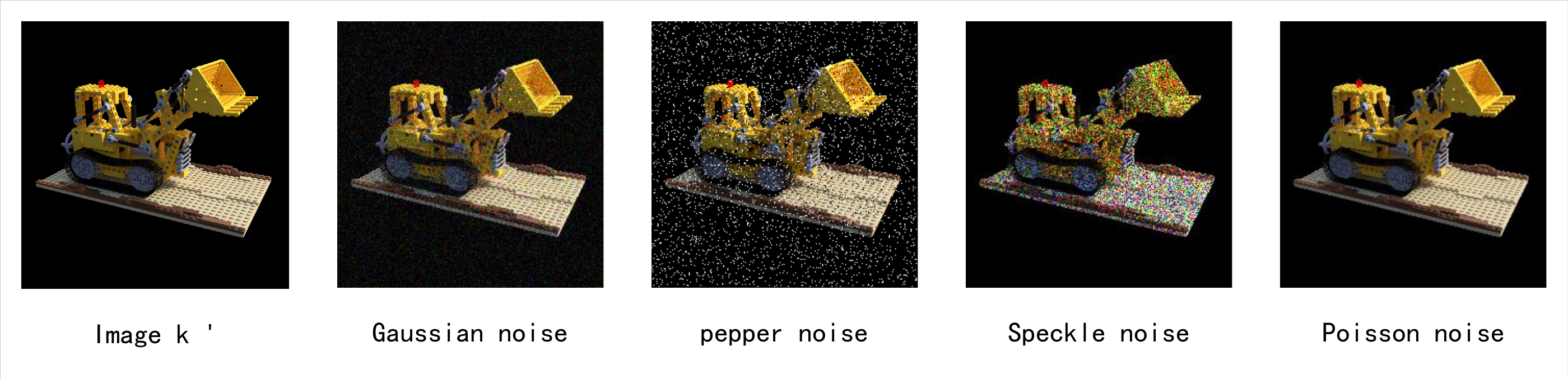}}
\caption{Example of noise layer.}
\end{figure}
\subsection{Neural radiation field}
The neural radiation field is a methodology employed for the generation of 3D scenes, entailing the utilization of a neural network model to establish a mapping relationship between pixel space and scene space. This mapping enables the prediction of radiation transmission within the scene. In this study, NeRF is trained using camera parameters associated with a confidential angle, facilitating the generation of a clandestine view image.
\begin{enumerate}
\item[1)]{\it NeRF network structure:} The NeRF model takes as input the 3D coordinate position, denoted as x=(x, y, z), and the direction, represented as d=($\theta$ , $\phi$), of the point in the input space. The output space comprises the color c=(r, g, b) of the corresponding point and the density $\theta$ of the corresponding voxel position.In the specific implementation, the position information of x and d is firstly encoded, followed by the input of x into the MLP network. This results in the output of $\theta$ and a 256-dimensional intermediate feature. The intermediate feature and d are then inputted into a fully connected layer to predict the color. Finally, 2D images are generated through volume rendering, which is a process transforming 3D data into 2D representation.Volume rendering involves obtaining the final pixel value of a 2D image by generating pixel samples along a ray in the viewing direction and performing a weighted sum of the pixel values c and volume density $\theta$ obtained from the 3D points reconstructed by 3D reconstruction. This process is expressed mathematically as shown in formula (1).
\begin{equation}
\begin{array}{l}
C(r) = \int_{{t_n}}^{{t_f}} {T(t)\sigma (r(t))c(r(t),d)dt,} \\
whereT(t) = \exp ( - \int_{{t_n}}^{{t_f}} {\sigma (r(s))} ds)
\end{array}
\end{equation}
Here, a ray, denoted as $\ {r(t)}$, is defined as $\ {r(t)=o+td}$, where (o) represents the photocenter position of the camera and (d) represents the directional vector. The cumulative transmittance of the ray along the path from the near end $\ {t_n}$ to the far end boundary $\ {t_f}$ is denoted as $\ {T(t)}$ .During the training process, the rendering of the overall scene is achieved by obtaining the image pixels and comparing them with the original perspective using the mean square error, also known as $\ {\cal L}_2$ loss or quadratic loss $\ {\cal L}_2$. The network is optimized based on this loss function. Additionally, a hierarchical representation rendering approach is adopted to improve rendering efficiency. This involves the simultaneous optimization of both the coarse network $\ {C_c}(r)$ and the fine network $\ {C_f}(r)$. By doing so, the weight distribution of the coarse network is utilized to better allocate samples in the fine network.
\begin{equation}
\begin{array}{l}
L = \sum\nolimits_{r \in R} {\left[ {\left\| {{{\hat C}_c}(r) - C(r)} \right\|_2^2 + \left\| {{{\hat C}_f}(r) - C(r)} \right\|_2^2} \right]} 
\end{array}
\end{equation}
In formula (2), R represents all the rays in the input view,$\ {C(r)}$ represents the actual pixel value of the input view, and by $\hat C(r)$  represents the predicted pixel value.

\item[2)]{\it Camera parameter:} NeRF takes a finite set of discrete images and camera parameters corresponding to the viewing angle to generate a continuous static 3D scene. It is capable of generating a new view image from an infinite number of angles. In this study, leveraging this feature of NeRF, we select the camera parameter (m) for a specific angle as the key. Since there are infinitely many angle options, the key space is infinite, ensuring the algorithm’s security. Camera parameters are categorized into internal parameters and external parameters. The external parameters determine the camera’s position and orientation, and their role is to transform points from the camera coordinate system to the world coordinate system. The camera parameter determines the projection property by mapping 3D coordinates in the camera coordinate system to the 2D image plane. In this study, we select camera parameters as the key to obtain the secret view image, ensuring image uniqueness. Moreover, the camera parameters consist only of two matrices, resulting in small key information.
\end{enumerate}
\subsection{Extraction network} 
The extraction network employs an over-parameterized approach to extract watermark information. Similar to the embedded network, the extraction network incorporates the concept of dense connections to enhance the transfer of features.
\begin{enumerate}
\item[1)]{\it overparameterization:}
Neural networks often encounter overfitting issues due to the abundance of parameters. However, this paper leverages this drawback as an advantage. Specifically, an extraction network with a substantial number of parameters is employed, where the secret view image and the original watermark image serve as inputs, while the output is the watermark image. Notably, when the secret-angle image is provided as an input, the watermark image can be accurately obtained. Conversely, if any other angle image is provided, the watermark image cannot be retrieved, thereby ensuring the algorithm’s security within this study.
\item[2)]{\it Extract network structure:} 
The network takes in an image s of size 3×W×H, which is then extracted and processed through a convolutional layer with an output channel of 32 to obtain the feature map E. Subsequently, the feature map F is obtained through an additional convolution on E. By concatenating E, F, and G and performing another convolution, the feature map G is obtained, resulting in a clear watermark image w’. The basic procedure is depicted in formulas (8) to (11).
\begin{equation}
\begin{array}{l}
\left\{ {\begin{array}{*{20}{c}}
{E = Con{v_{3 -  > 32}}(s)}\\
{F = Con{v_{32 -  > 32}}(E)}\\
{G = Con{v_{32 + 32 -  > 32}}(Cat(E,F))}\\
{w' = Con{v_{32 + 32 + 32 -  > D}}(Cat(E,F,G))}
\end{array}} \right.
\end{array}
\end{equation}
The convolutional kernel settings of the extraction network remain consistent with those of the embedded network, and a 3×W×H watermark image is outputted after the final convolutional layer. Once the extraction network is trained, the model parameters are saved. Importantly, the watermark image cannot be extracted from input images generated by other NeRF perspectives. For a comprehensive illustration of the extraction network’s architecture, refer to Figure 6.
\begin{figure}
\centerline{\includegraphics[width=\columnwidth]{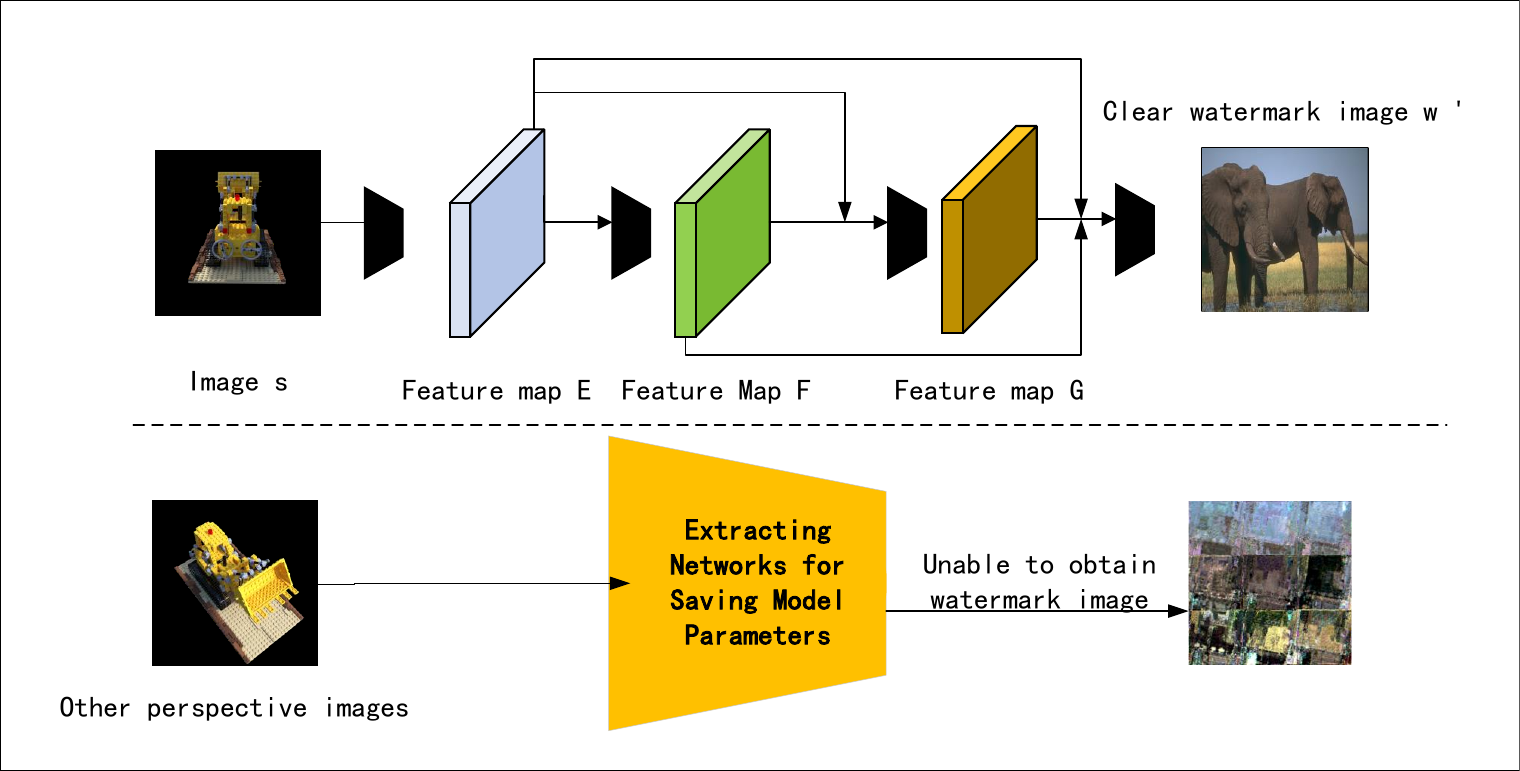}}
\caption{W atermark Extraction Network Structure.}
\end{figure}
\end{enumerate}
\subsection{Loss function} 
The loss function primarily comprises two components: carrier image content $\ {\cal L}e$ and watermark image content $\ {\cal L}d$. $\ {\cal L}e$ represents the content loss of the carrier image, which is calculated as the Mean Square Error (MSE) between the carrier image and the image containing the watermark, as specified in Equation (12).
\begin{equation}
\begin{array}{l}
{\cal L}e = \alpha \frac{1}{{3 \times H \times W}}||k - k'||_2^2
\end{array}
\end{equation}
Here, $\alpha$ denotes the weight parameter, k represents the carrier image, and k’ represents the image containing the watermark. The watermark image content loss, denoted as $\ {\cal L}d$, is defined by formula (13).
\begin{equation}
\begin{array}{l}
{\cal L}d = \beta \frac{1}{{1 \times H \times W}}||w - w'||_2^2 + \gamma (1 - {\cal L}_{ssim}) + \mu (1 - {\cal L}_{msssim})
\end{array}
\end{equation}

In the equation, $\beta$, $\gamma$, and $\mu$ represent the weight parameters, w denotes the original watermark image, and w’ denotes the extracted watermark image.$\ {\cal L}_{{\rm{ssim}}}$ refers to the loss evaluated by Structural SIMilarity (SSIM), while $\ {\cal L}_{{\rm{msssim}}}$ corresponds to the loss evaluated by Multi-Scale Structural SIMilarity (MS-SSIM [26]). The computation formulas for SSIM and MS-SSIM are presented in formulas (14) and (15), respectively.
\begin{equation}
\begin{array}{l}
SSIM(x,y) = \frac{{(2{\mu _x}{\mu _y} + {C_1})(2{\sigma _{xy}} + {C_2})}}{{({\mu _x}^2 + {\mu _y}^2 + {C_1})({\sigma _x}^2 + {\sigma _y}^2 + {C_2})}}
\end{array}
\end{equation}
\begin{equation}
\begin{array}{l}
MS - SSIM(x,y) = \prod\limits_{{\rm{m}} = 1}^M {{{(\frac{{2{\mu _x}{\mu _y} + {C_1}}}{{\mu _x^2 + \mu _y^2 + {C_1}}})}^{{\beta _m}}}} {(\frac{{2{\sigma _{xy}} + {C_2}}}{{\sigma _x^2 + \sigma _y^2 + {C_2}}})^{{\gamma _m}}}
\end{array}
\end{equation}
In the aforementioned equation, $\ {\mu _x}$ and $\ {\sigma _x}$ denote the mean and variance of image x, while $\ {\mu _y}$ and $\ {\sigma _y}$ represent the mean and variance of image y, respectively. Additionally, $\ {\sigma _{xy}}$ denotes the covariance of images X and Y. The variable M symbolizes the different scales, and $\ {\beta _m}$ and $\ {\gamma _m}$ represent the relative importance between images. For this study, $\ {\beta _m}$ and $\ {\gamma _m}$ are both set to 1. Constants C1 and C2 are defined, with a value of k1 = 0.01 and k2 = 0.03, respectively. Notably, L refers to the dynamic range of pixel values, with a specified value of 255.
\section{Experimental results and analysis}

\subsection{Experimental setup}
For this experiment, the NeRF model is trained using the NeRFSynthetic dataset, which includes images of Lego, Chair, Ship, and other objects, while the ImageNet dataset is utilized for selecting random images as watermark images for testing purposes. The NeRFSynthetic dataset encompasses images captured from various angles of the model, accompanied by the respective camera parameters. Notably, the ImageNet dataset is a considerable dataset comprising 14,197,122 images, of which 100 images are specifically chosen as watermark images for this study. The experiment was conducted on a computer running Windows 11 with a capacity of 64GB. The embedded network and extraction network were optimized using the Adam optimizer. The input image size was uniformly adjusted to 256×256 pixels, with a learning rate of 0.0001 and a total of 50,000 epochs. The parameters of the loss function were set as follows: $\ \alpha $ = 0.3, $\ \beta $ = 0.3, $\ \gamma $ = 0.5, and $\ \mu $ = 0.5.
\subsection{imperceptibility}
Imperceptibility, also referred to as perceptual transparency, implies that there should be no perceivable difference between the embedded watermark image and the original image. In general, Mean Squared Error (MSE) is employed to evaluate the quality of image alterations, Peak Signal-to-Noise Ratio (PSNR) is used to assess the compression quality of the image, and Structural SIMilarity (SSIM) measures the similarity between the embedded watermark image and the original image. Additionally, LPIPS [27] was employed to evaluate the perception loss and image quality during the evaluation process. PSNR is calculated based on MSE. A higher PSNR and SSIM are indicative of better quality, while a lower LPIPS is desirable. The calculation formulas for MSE and PSNR are presented in equations (16) and (17), respectively.
\begin{equation}
 \begin{array}{l}
  MSE = \frac{1}{{W \times H}}{\sum\limits_{i = 1}^W {\sum\limits_{j = 1}^H {({X_{i,j}} - {Y_{i,j}})} } ^2}
  \end{array}
  \end{equation}
\begin{equation}
\begin{array}{l}
PSNR = 20 \times \lg (sc) - 10 \times \lg (MSE)
\end{array}
\end{equation}
In the context, X and Y represent two images with dimensions of W×H, respectively. The variable sc indicates the scaling factor, which is typically assigned a value of 2.

\begin{table*}[!htbp]
		\caption{Quantitative analysis of imperceptibility} 
		\centering
		\begin{tabular}{*{8}{c}} 
        \hline
			\toprule 
            \multicolumn{2}{c}{\textbf{Data set }} &
            \multicolumn{3}{c}{\textbf{ Embedded network}}&            
		    \multicolumn{3}{c}{\textbf{Extract network}} \\  
			Embedd & Extract &  PSNR(dB)↑ & SSIM↑& LPIPS↓&PSNR(dB)↑ &  SSIM↑&  LPIPS↓\\
   \hline 
            lego & ImageNet & 35.62 & 0.9451 & 0.1523  & 34.81 & 0.9378& 0.0988 \\
			\midrule 
			chair & ImageNet & 33.41 & 0.9432  & 0.1746&34.35 & 0.9389  & 0.1124 \\
           \midrule 
			ship & ImageNet & 32.76 & 0.9421 &0.1279  &   35.21 & 0.9353 & 0.0992 \\
             \midrule 
             drums & ImageNet & 33.83 &0.9476  &0.1456  &33.62  & 0.9195 & 0.1326 \\
              \midrule 
              hotdog & ImageNet & 34.56 &0.9473&  0.0994 &33.19 & 0.9468   & 0.1258 \\
               \midrule 
               materials & ImageNet & 32.41 &  0.9351 & 0.1348 &31.46 & 0.9234 & 0.1127\\ 
			\bottomrule 
               \hline
		\end{tabular}
\end{table*}

\begin{table}[!htbp]
		\caption{Quantitative analysis of watermark extraction effect} 
		\centering
		\begin{tabular}{*{7}{c}}  
			\toprule 
			Angular change & 0 & 1 & 3 & 7 & 10 & 15 \\\hline 
            PSNR & 30.75 & 25.40 & 21.70 & 19.68 & 18.70 & 17.95\\  
			\midrule 
		    SSIM & 0.9137 & 0.7243 & 0.6609 & 0.6303 & 0.6163 & 0.5988 \\ 
           \midrule 
		    LPIPS & 0.1886 & 0.2721 & 0.3103 & 0.3436 & 0.3565 & 0.3684  \\
			\bottomrule 
               \hline 
		\end{tabular}
        \begin{tabular}{*{7}{c}}  
			\toprule 
			Angular change  & 30 & 60 & 90 & 180 & 300 & 340 \\\hline 
            PSNR  & 16.39 & 15.30 & 15.59 & 16.55 & 15.92 & 17.10 \\ 
			\midrule 
			SSIM  & 0.5490 & 0.5238 & 0.5372 & 0.5706 & 0.5379 & 0.5803 \\
           \midrule 
			LPIPS & 0.4030 & 0.4230 & 0.4118 & 0.3961 & 0.4078 & 0.3778 \\
			\bottomrule 
              \hline 
		\end{tabular}
\end{table}

\begin{enumerate}

\item[1)]{\it Image quality:} The experimental results of this algorithm are presented in the table. The PSNR, SSIM, and LPIPS values reported in the table for both the embedded networks and the extraction networks represent their respective mean experimental results. As evident from the table, the embedded network and extraction network exhibit high PSNR and SSIM values, while demonstrating low LPIPS values. These findings indicate that the algorithm proposed in this paper generates embedded watermark images that possess high imperceptibility, as reflected by the extracted watermark images.
\item[2)]{\it Image effect:}In this experiment, an image was embedded in each of the Lego, Chair, Ship, Drums, Hotdog, and Materials datasets. Subsequently, these images, along with others, underwent training using the NeRF algorithm. Following the training, watermark information was then extracted from the rendered new visual images. The visual outcomes of the embedded watermark image and the extracted watermark image are depicted in Figure 7 and Figure 8, respectively.

\begin{figure}
\centerline{\includegraphics[width=\columnwidth]{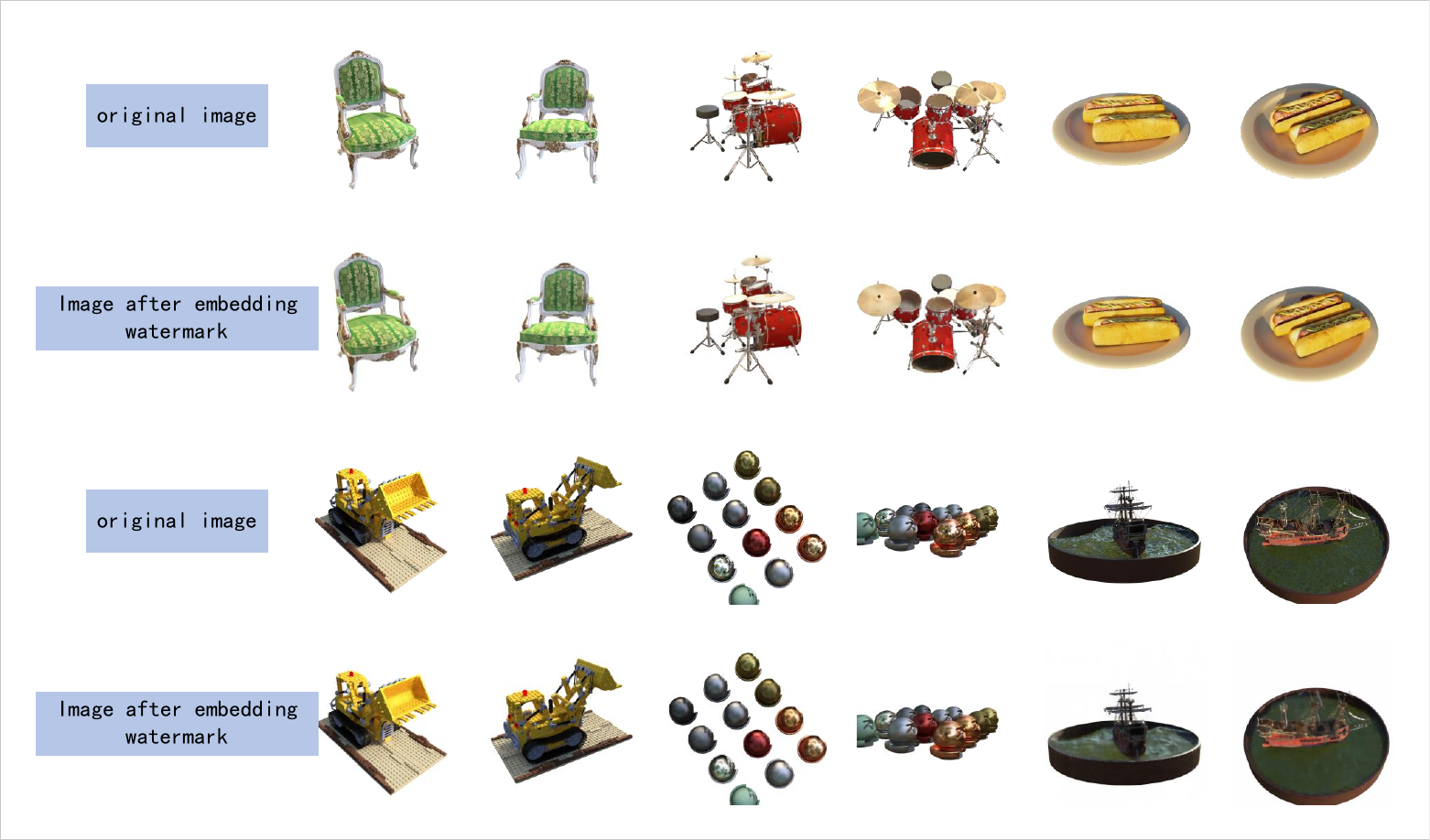}}
\caption{Visual effect after embedding watermark.}
\end{figure}

\begin{figure}
\centerline{\includegraphics[width=\columnwidth]{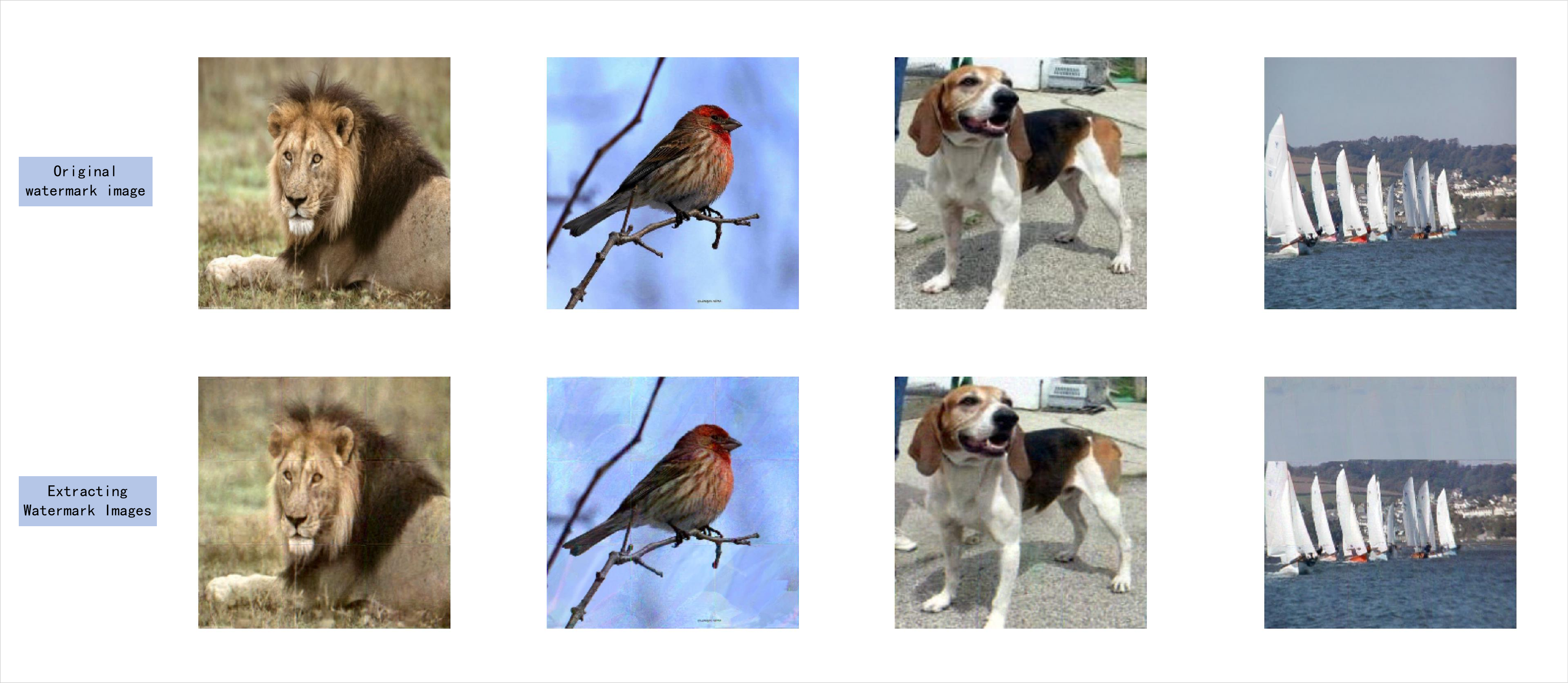}}
\caption{Visual effects of extracting watermark images.}
\end{figure}

\end{enumerate}

\subsection{robustness}
Robustness refers to the watermark’s ability to withstand attacks. In this study, the process of neural radiation field being attacked was simulated by introducing noise to the images in the training set. The robustness of the watermark is typically assessed using metrics such as Bit Error Rate (BER) and Normalized Correlation (NC).
\begin{enumerate}
\item[1)]{\it Bit error rate:}The bit error rate (BER) is defined as the reciprocal of the peak signal-to-noise ratio (PSNR), with its formula presented in equation (18).
\begin{equation}
\begin{array}{l}
BER = \frac{1}{{PSNR}}
\end{array}
\end{equation}
The bit error rate (BER) quantifies the extent to which the original image is altered during the watermarking process. In this experiment, noise was introduced to the image from the Lego dataset, while the lion image from the ImageNet dataset was utilized as the watermark image to obtain the experimental results. The BER values associated with the extracted watermark under various noise attacks are presented in Table 2.
\begin{table}
\caption{Error rate}

\centering
\setlength{\tabcolsep}{3pt}
\begin{tabular}{ccccc}
\hline noise type & Gaussian noise & Pepper Noise & Speckle noise & Poisson noise \\
\hline BER & 0.02964 & 0.03081 & 0.03185 & 0.02836 \\
\hline
\end{tabular}
\label{tab1e}
\end{table}
\item[2)]{\it Normalized correlation coefficient:} To quantitatively assess the similarity between the extracted watermark image and the original watermark image, the normalization coefficient is employed as the evaluation criterion. Its definition is presented in equation (19).
\begin{equation}
\begin{array}{l}
NC(w,w') = \frac{{\sum\limits_{x = 1}^M {\sum\limits_{y = 1}^N {w(x,y)w'(x,y)} } }}{{\sqrt {\sum\limits_{x = 1}^M {\sum\limits_{y = 1}^N {w{{(x,y)}^2}} } } \sqrt {\sum\limits_{x = 1}^M {\sum\limits_{y = 1}^N {w'{{(x,y)}^2}} } } }}
\end{array}
\end{equation}
Here, w(x,y) refers to the original watermark image, w’(x,y) represents the extracted watermark image, M and N denote the image resolution. The normalized correlation coefficient (NC) signifies the similarity between the original image and the extracted watermark image, ranging from 0 to 1. A higher value closer to 1 indicates better robustness.In this experiment, varied noise types were employed to attack the watermark-embedded image, followed by the extraction of the watermark image and calculation of the corresponding NC. The normalized correlation coefficient of the extracted watermark image following the attack is depicted in Figure 9.
\begin{figure}
\centerline{\includegraphics[width=\columnwidth]{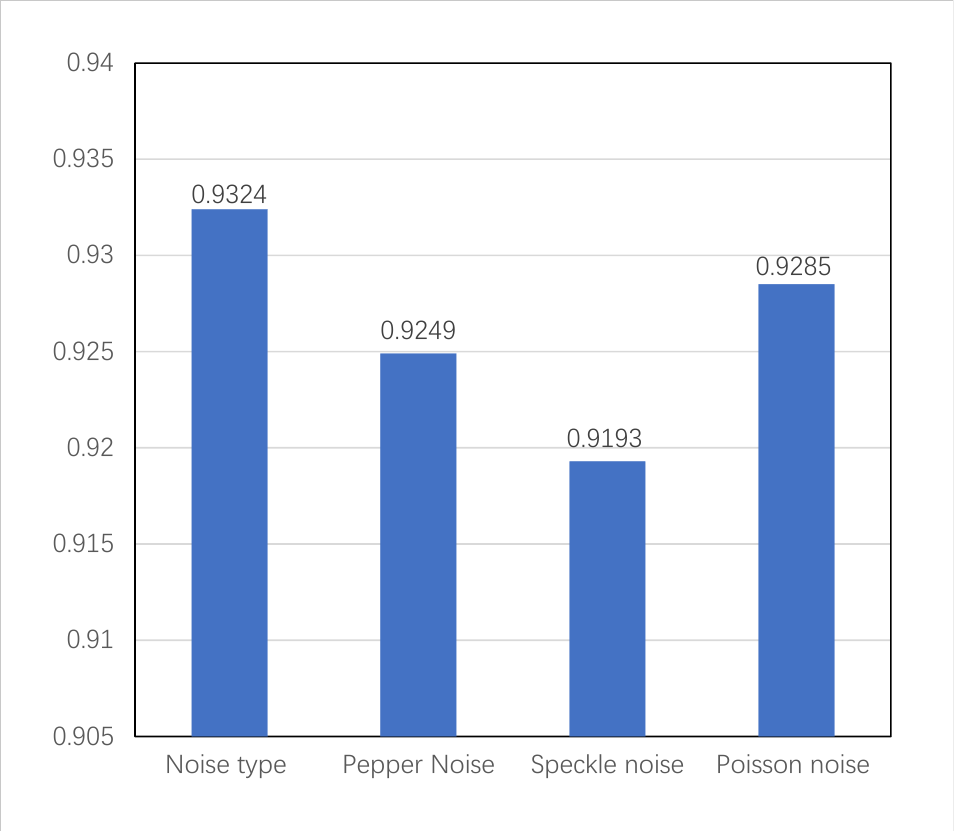}}
\caption{Normalized correlation coefficient graph.}
\end{figure}
Upon analyzing the findings presented in the figure, it is evident that the proposed watermarking scheme in this study exhibits commendable anti-interference capability.
\end{enumerate}

\subsection{The effectiveness of the secret perspective}
To evaluate the efficacy of the hidden perspective, this study investigates the impact of images generated from different viewpoints on the watermark extraction performance. For this experiment, a specific image from the Lego dataset was selected as the host image for watermark embedding, while an elephant image from the image dataset was chosen as the watermark image. Images rendered using the NeRF technique under various perspectives were inputted into the extractor for evaluation. The experimental outcomes are depicted in Figure 10.
\begin{figure}
\centerline{\includegraphics[width=\columnwidth]{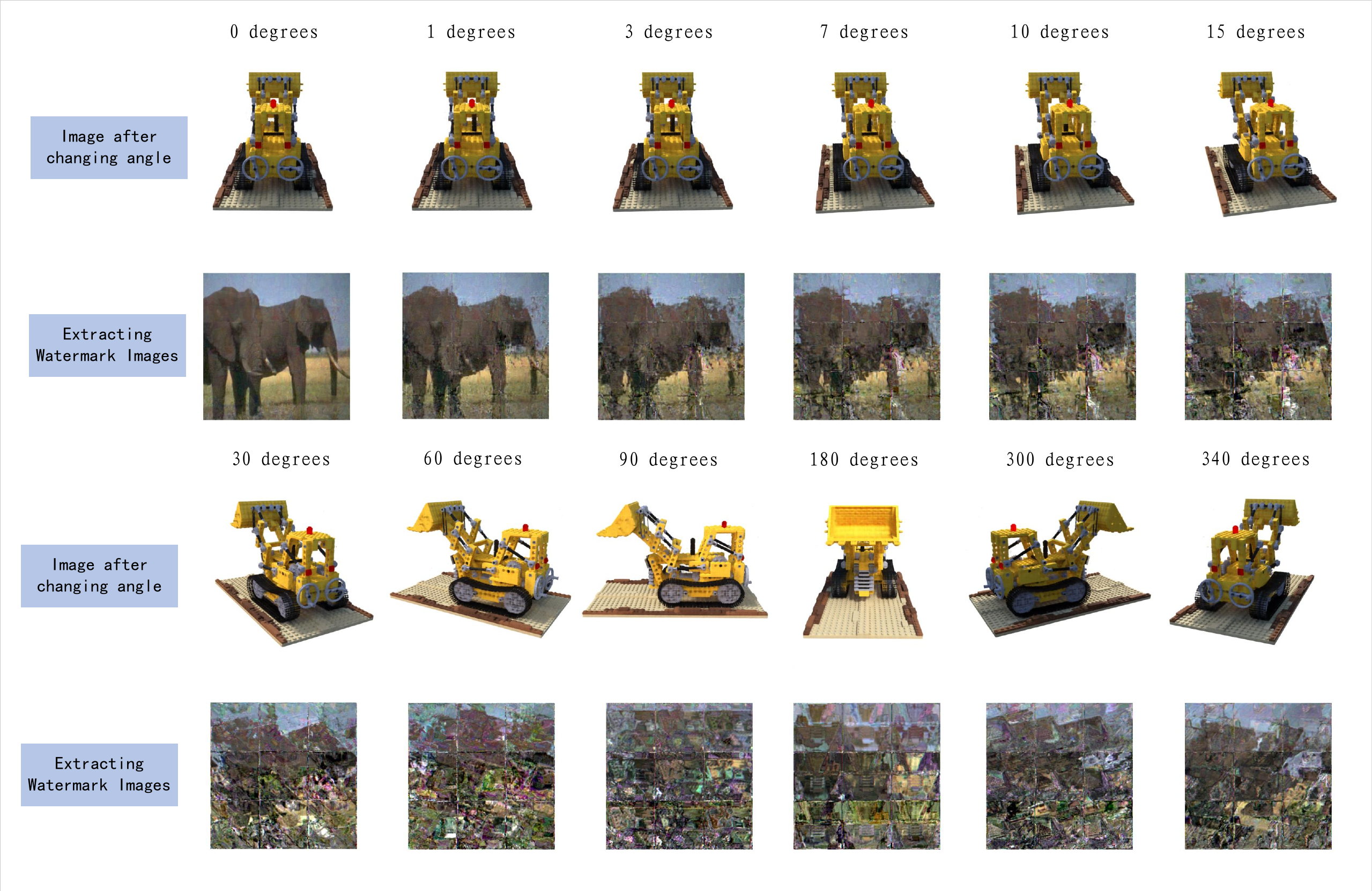}}
\caption{Comparison of watermark extraction effects on different visual images.}
\end{figure}

The experimental results indicate that as the rotation angle increases, the extracted watermark image progressively becomes more blurred, eventually reaching a point where the watermark image extraction is no longer feasible. To quantitatively analyze the impact of the rotation angle on watermark extraction, the values of PSNR, SSIM, and LPIPS are also provided, representing the comparison between the extracted watermark and the original image. These results are presented in the table. The angle variation specified in the table corresponds to the rotation angle around the central z-axis.

Based on the experimental data, it is observed that smaller rotation angles yield higher PSNR and SSIM values for the extracted watermark, demonstrating superior similarity to the original watermark image. Additionally, the lower LPIPS values further support the notion of the proposed algorithm’s efficacy in safeguarding model copyright.

\section{Conclusion}
This paper introduces the novel concept of neural implicit representation watermarking, deploying neural network watermarking technology to protect neural networks with implicit data representations. A watermarking algorithm specific to neural radiation fields is devised. Initially, the embedded network is employed to imbue watermarking onto images within the training dataset. Then, NeRF is utilized to generate 3D models and render images from novel perspectives, thereby enabling the watermark extraction process. Experimental results demonstrate the algorithm’s security, as any viewing angle within the infinite viewing angle space is considered a key. The watermark extraction not only manifests superior visual quality, but also exhibits robustness. In future work, enhancements to the extraction network structure can be pursued to diminish the likelihood of extracting watermark information from adjacent view images, further fortifying the algorithm’s security.

\bibliographystyle{IEEEtran}

\end{document}